\documentstyle[12pt]{article}

\input epsf

\def\beq{\begin{equation}}
\def\eeq{\end{equation}}

\def\al{\alpha}
\def\bt{\beta}
\def\Ga{\Gamma}
\def\de{\delta}

\def\lam{\lambda}

\def\l{\left (}
\def\r{\right )}
\def\fr{\frac}
\def\la{\label}
\def\hs{\hspace}
\def\vs{\vspace}

\def\ov{\overline}
\def\ran{\rangle}
\def\lan{\langle}
\def\ti{\tilde}

\begin{document}

\begin{center}
{\Large \bf  Low Scale Theories: Light Neutrinos \\
and Unification of Gauge Couplings
}
\end{center}
\vspace{0.5cm}
\begin{center}
{\large Zurab Tavartkiladze}
\vspace{0.5cm}

{\em E. Andronikashvili Institute of Physics\\
380077 Tbilisi, Georgia\\
E-mail: z\_tavart@osgf.ge
}
\end{center}

\begin{abstract}
Mechanism for generation of suppressed neutrino masses, within low
scale theories, is considered. The mechanism do not have extradimensional
nature and is realized through extended
$SU(2)_L$ scalar multiplets. 
Latters, in some cases, are also crucial for successful low
scale unification. 
\end{abstract}


\hs{-0.6cm}1.~{\it Introduction.}~~Despite great success in solving the
gauge
hierarchy problem,
within low scale theories \cite{tevgr1}, there are
problems and issues which should be reconsidered from a new viewpoint.
One of the actual task is to understand the suppression of
neutrino masses, which due
to low fundamental scale, expected to be unacceptably large.
In \cite{bulknu1}, for generation of suppressed neutrino masses,
existence of extra dimensions have played crucial role. 
In \cite{neutsup}, for the same purpose, 
together with right handed neutrinos was introduced
additional scalar doublet with a sufficiently tiny VEV.

Here we demonstrate how the suppressed neutrino masses can be generated
through the extended charged $SU(2)_L$ scalar multiplets \cite{mypaper}.
Namely, $4$, $5$ or $6$ dimensional plets should be applied
respectively, depending on a value of fundamental scale $M_f$. These
multiplets are crucial also for non SUSY low scale unification. For
SUSY scenarios, some new possibilities of low scale unification are also
found.


\vspace{0.2cm}

\hs{-0.6cm}2.~{\it Generation of Suppressed Neutrino Masses.}~~Introduce
$\Phi $ scalar
in  $(4,~-3)$ representation (REP) of  $SU(2)_L\times U(1)_Y$.
In this $U(1)_Y$ normalization, $Y(l)=1$
($l$ is lepton doublet). For avoiding
$(lh^{+})^2/M_f$ type operators ($h$ is SM Higgs
doublet), we assume that in fermion sector lepton number $L$ is conserved. 
Prescribing
to $\Phi $ lepton number $-2$, the Yukawa couplings responsible for
neutrino masses possess $U(1)_L$ symmetry
\vs{-0.2cm}
\beq
{\cal L}_{\nu }=\hat{\lam }_{\nu }ll \Phi h/M_f +h.c.~,
\la{4nulag}
\eeq
where $\hat{\lam }_{\nu }$ is matrix in a family space. 
{}For low $M_f$, scalar $\Phi $ should develop
tiny VEV along its neutral component in order to
generate suppressed neutrino masses. This is naturally
insured
through the scalar potential of $h$ and $\Phi $ fields:
\vs{-0.2cm}
$$
{\cal V}(h, \Phi )=\fr{\lam_h}{2}\l h^{+}h-m^2\r^2+
\fr{\lam_{\Phi }}{2}\l \Phi^{+}\Phi +M^2\r^2+
$$
\vspace{-0.7cm}
\beq
\lam_1\l \Phi^{+}\Phi \r \l h^{+}h \r +
\lam_2\l \Phi^{+}h \r \l\Phi h^{+} \r -
\lam \l \Phi h^{3}+\Phi^{+}(h^{+})^3\r ~,
\la{4pot}
\eeq
where $m$ is Higgs doublet mass $\sim 100$~GeV.
Last term in (\ref{4pot})
mildly violates $U(1)_L$ and for all positive parameters in (\ref{4pot})
system will have global minimum with non zero $\lan \Phi \ran $.
The extremum conditions for (\ref{4pot}) will be:
\vs{-0.1cm}
$$
\lam_h(v^2-m^2)+(\lam _1+\lam _2)V^2-3\lam Vv=0~,
$$
\vspace{-0.8cm}
\beq  
\lam_{\Phi }(V^{2}+M^{2})V+(\lam_1+\lam_2)Vv^2-\lam v^3~=0~,
\la{min}
\eeq
and {}for
$\lam_{\Phi }M^{2}\gg (\lam_1+\lam_2)m^2~$, one can easily obtain
\vspace{-0.1cm}
\beq
v=m+{\cal O}\l m^3/M^2\r ~, ~~~~
V=\lam v^3/(\lam_{\Phi }M^2) +{\cal O}\l m^5/M^4\r ~.
\la{4sol}
\eeq  
Note, that although the mass of $\Phi $ is much larger than $v$, the
hierarchy is not destabilized, because $\Phi$'s VEV in (\ref{4sol}) is   
tiny and quartic terms in (\ref{4pot}) practically do not affect $v$.
Using (\ref{4sol}) in (\ref{4nulag}), for neutrino masses we will have
\vspace{-0.1cm}
\beq
\hat{m}_{\nu }=\hat{\lam }_{\nu }vV/M_f\simeq 
\lam \hat{\lam }_{\nu }v^4/(\lam_{\Phi }M^2M_f),
\la{numas4}
\eeq
and desirable value $\hat{m}_{\nu }=(1-4\cdot 10^{-2})$~eV is obtained
for $M\simeq M_f=(1-3)\cdot 10^{3}$~TeV with $v=174$~GeV,
$ \lam \hat{\lam }_{\nu }/\lam _{\Phi }\sim 1 $.

If we wish to build scenario with lower $M_f$, higher
$SU(2)_L$ REPs must be introduced. Namely, if now $\Phi $ is
$5$-plet of $SU(2)_L$ with $Y(\Phi )=-4$, then instead of (\ref{4nulag})
we will have
${\cal L}_{\nu }=\hat{\lam }_{\nu }ll \Phi h^{2}/M_f^{2}+
{\rm h.c.}$~,
and in potential (\ref{4pot})
last term will be replaced with
$-\lam' (\Phi h^{4}+\Phi ^{+}{h^{+}}^{4})/M_f $.
For this case it is easy to verify that
$v\simeq m~,~
V\simeq \lam ' v^4/(\lam_{\Phi }M^2M_f)$ and
consequently for neutrino masses
\vspace{-0.3cm}
\beq
\hat{m}_{\nu }=\hat{\lam }_{\nu }v^2V/M_f^2\simeq
\lam' \hat{\lam }_{\nu }v^6/(\lam_{\Phi }M^2M_f^3)~,
\la{numas5}
\eeq
which for $\hat{m}_{\nu }=(1-0.1)$~eV,
$\lam' \hat{\lam }_{\nu }/\lam _{\Phi }\sim 1 $
require relatively low scales  $M\simeq M_f=(30-50)$~TeV.

Scales $M_f$, $M$ can be easily reduced even down to few TeV, if
$\Phi $ belongs to $(6,~-5)$ REP of $SU(2)_L\times
U(1)_Y$.
Then instead the last term in (\ref{4pot}) we will have
$-\lam'' (\Phi h^{5}+\Phi ^{+}{h^{+}}^{5})/M_f^2 $
and relevant Yukawa couplings will be
$\hat{\lam }_{\nu }ll \Phi h^{3}/M_f^{3}$. By simple analyses one
can
easily obtain that in this case
\vspace{-0.2cm}
\beq
\hat{m}_{\nu }\simeq
\lam'' \hat{\lam }_{\nu }v^8/(\lam_{\Phi }M^2M_f^5)~,
\la{numas6}
\eeq
and $(1-0.1)$~eV neutrino masses
are generated for $M\simeq M_f=(7-10)$~TeV.

Supersymmetrizing these scenarios, together with superfield $\Phi $
(which denote $4$, $5$ or $6$-plets) we introduce its conjugate
superfield $\ov{\Phi } $. Relevant superpotential is
\vspace{-0.3cm}
\beq
W_{\Phi }=M\ov{\Phi } \Phi-
\l \lam_{\Phi d}\Phi h_d^{3+n}+
\lam_{\Phi u}\ov{\Phi } h_u^{3+n}\r/M_f^{1+n}~,
\la{sup}
\eeq
where $n=0, 1, 2$ for scenarios with $\Phi+\ov{\Phi }$   
in $4$, $5$ and $6$ REPs of $SU(2)_L$ respectively.
$h_u$, $h_d$ denote doublet-untidoublet pair of MSSM and
$\lam_{\Phi d}$, $\lam_{\Phi u}$ are positive dimensionless couplings
of the order of one.
Yukawa superpotential, responsible for neutrino masses, will be
\vspace{-0.2cm}
\beq
W_{\nu }=\hat{\lam}_{\nu }ll\Phi h_d^{n+1}/M_f^{n+1}~.
\la{yuksup}
\eeq
After that SUSY and EW symmetry breaking take place,
non zero $\lan h_u \ran,~\lan
h_d\ran $
are generated and from (\ref{sup}) one can easily verify
$\lan \Phi \ran \simeq \lam_{\Phi u}\lan h_u\ran^{n+3}/(MM_f^{n+1})$. 
Using this and also (\ref{yuksup}), we will get
\vspace{-0.2cm}
\beq
\hat{m}_{\nu }=\hat{\lam }_{\nu }\lam_{\Phi u}
\sin^{n+3}\bt \cos^{n+1}\bt \cdot
v^{2n+4}/(MM_f^{2n+2})~,
\la{susynumas}
\eeq
where we have used $\lan h_u\ran=v\sin \bt $,
$\lan h_d\ran=v\cos \bt $. 
{}For
$v=174$~GeV, $\tan \bt \simeq 1$, neutrino masses
$m_{\nu } \sim  (1-0.1)$~eV are obtained within various
scenarios:
\vspace{-0.1cm}
\begin{equation}
M\simeq M_f=\left\{ \begin{array}{lll}
(0.6-1.3)\cdot 10^3~{\rm TeV}; & n=0,~ {\rm case~ with
~4-plets} \\
(20-30)~{\rm TeV}; & n=1,~ {\rm case~ with~ 5-plets} \\
(4.7-6.5)~{\rm TeV}; & n=2,~ {\rm case~ with~ 6-plets}
\end{array}
\right.~.
\la{susyrange}
\end{equation}
Larger values of $\tan \bt $ would give stronger suppression for 
$\hat{m}_{\nu}$
in (\ref{susynumas}), giving possibility to reduce mass scales in 
(\ref{susyrange}) by few factors.

Mechanisms which we have suggested here, provide adequate suppressions of
neutrino masses and this suppressions occur through proper choice of
$\Phi $ scalar in appropriate $SU(2)_L\times U(1)_Y$ REP.
Neutrino mass scale $\sim (0.1-1)$~eV is natural for atmospheric
anomaly if three family neutrinos are either hierarchical in mass or
degenerate, respectively. For simultaneous accommodation of atmospheric and
solar neutrino data, one can introduce flavor symmetries and build
different oscillation scenarios in a spirit of \cite{osc}.

\vspace{0.2cm}

\hs{-0.6cm}3.~{\it Gauge Coupling Unification.}~~Due to extra spacelike
dimensions (with radius $R$), gauge couplings can get power low
runnings starting from scale $\mu_o=1/R$. This gives
possibility for low scale unification \cite{uni1}. Solutions of 1-loop
RGEs are 
\vspace{-0.3cm}
\beq
\al^{-1}_G=\al_a^{-1}(M_Z)- \fr{b_a}{2\pi }\ln \fr{M_G}{M_Z}-
\fr{\ti b^i_{a}}{2\pi }\ln \fr{M_{G}}{M_i}-
\fr{\hat{b}^i_{a}}{2\pi }P_{\de }^{(\mu_i)}~,
\la{alps}
\eeq
where $\al _{1,2,3}$ denote gauge couplings 
of $U(1)$, $SU(2)_{L}$ and $SU(3)_{c}$ respectively,   
$b_{a}$ is b-factors of SM/MSSM, $\ti{b}_a^i$ come from contribution of
additional states of mass $M_i$($<\hs{-0.1cm}M_G$ GUT scale$\simeq M_f$),
$\hat{b}_a^i$ come from Kaluza-Klein (KK) states;
\vspace{-0.4cm}
\beq
P_{\de }^{(\mu_i)}=\fr{X_{\de }}{\de } \left [\l
M_G/\mu_i\r^{\de }\hs{-0.3cm}-1\right ]\hs{-0.1cm}-
\ln M_G/\mu_i~,~~
X_{\de }=\fr{2\pi ^{\de /2}}{\de \Ga (\de /2)}~,~~
\mu _i^2=M_i^2+\mu_0^2~.
\la{deKK}
\eeq
{}From (\ref{alps}), (\ref{deKK}) one can see that
for various scenarios successful unification
with $\al_s\simeq 0.119$ is achieved for:
{\bf a)}~Non SUSY scenario with two $\Phi (4)$-plets and
values of extra dimensions and scales:
\vspace{-0.1cm}
$$
(\de ~, ~M_G/\mu_0~,~M_G/M~,~M_G)=
(1,~~~9.78,~~~6.51,~~~10^{3.51}~{\rm TeV})~,
$$
\vspace{-0.7cm}
\beq
(2,~~~3.45,~~~2.83,~~~10^{3.26}~{\rm TeV}) ~,~   
(3,~~~2.36,~~~2.071,~~~10^{3.18}~{\rm TeV}) ~,\cdots
\la{sol4}
\eeq
{\bf b)}~non SUSY case with one $\Phi (5)$-plet and
\vspace{-0.2cm}
$$
(\de ~, ~M_G/\mu_0~,~M_G/M~,~M_G)=
(1,~~~11.55,~~~6.9,~~~10^{1.82}~{\rm TeV})~,
$$
\vspace{-0.7cm}
\beq
(2,~~3.735,~~2.92,~~10^{1.67}~{\rm TeV}) ~,~
(3,~~2.486,~~2.12,~~10^{1.61}~{\rm TeV}) ~,\cdots
\la{sol5}
\eeq
{\bf c)}~non SUSY case with one $\Phi (6)$-plet and 
\vspace{-0.2cm}
$$
(\de ~, ~M_G/\mu_0~,~M_G/M~,~M_G)=
(1,~~~12.25,~~~4.55,~~~10.7~{\rm TeV})~,
$$
\vspace{-0.7cm}
\beq
(2,~~~3.837,~~~2.35,~~~8.69~{\rm TeV})~,~
(3,~~2.486~,~2.12,~~~8.11~{\rm TeV}) ~,\cdots
\la{sol6}
\eeq
{}For {\bf a)}\hs{0.5mm}-\hs{0.5mm}{\bf c)} cases $\Phi $-plets are
crucial for unification.
Number of chiral families, with
KK excitations, can be $\eta =0\div 3$. For all this $\eta $, the $\al_G$
remain perturbative. 
SUSY unification require one additional $SU(3)_c$ adjoint
state (for each scenario)
with KK excitations and without zero mode wave function. 
Consequently, there are different cases of unification ($\al_s=0.119$):
{\bf d)}~SUSY scenario with one pair of $\Phi (4)+\ov{\Phi }(\ov 4)$
supermultiplets and
\vspace{-0.2cm}
$$
(\de ~, ~M_G/\mu_0~,~M_G/M~,~M_G)=
(1,~~~18.57,~~~10.56,~~~10^{3.13}~{\rm TeV}) ~,
$$
\vspace{-0.7cm}
\beq
(2,~~4.78,~~3.657,~~10^{2.96}~{\rm TeV}) ~,~
(3,~~2.937,~~2.465,~~10^{2.92}~{\rm TeV}) ~,\cdots
\la{sol4s}
\eeq
{\bf e)}~SUSY scenario with one pair of $\Phi (5)+\ov{\Phi }(\ov 5)$
supermultiplets and
\vspace{-0.2cm}
$$
(\de ~, ~M_G/\mu_0~,~M_G/M~,~M_G)=
(1,~~~18.02,~~~6.09,~~~10^{1.46}~{\rm TeV}) ~,
$$
\vspace{-0.7cm}
\beq
(2,~~4.683,~~2.742,~~10^{1.39}~{\rm TeV}) ~,~
(3,~~2.895,~~2.029,~~10^{1.36}~{\rm TeV}) ~,\cdots
\la{sol5s}
\eeq
{\bf f)}~SUSY scenario with one pair of $\Phi (6)+\ov{\Phi }(\ov 6)$
supermultiplets and
\vspace{-0.2cm}
$$
(\de ~, ~M_G/\mu_0~,~M_G/M~,~M_G)=
(1,~~~16.84,~~~3.9,~~~5.74~{\rm TeV}) ~,
$$
\vspace{-0.7cm}
\beq
(2,~~4.515,~~2.169,~~5.25~{\rm TeV}) ~,~
(3,~~2.823,~~1.729,~~5.12~{\rm TeV}) ~,\cdots
\la{sol6s}
\eeq
In cases {\bf d)}\hs{0.5mm}-\hs{0.5mm}{\bf f)} only $\eta =0$ is
allowed. For higher values of
$\eta $ gauge couplings become non perturbative.
Through analyses, we have taken $M_f\simeq M_G$. It is possible to have
$M_G$, by few factors and even more, below the $M_f$. This would reduce
scales $\mu_0$, $M$, making scenarios easily testable on a future
colliders.
\vspace{0.1cm}

\hs{-0.6cm}{\it Acknowledgements}~~It is pleasure to thank the organizers
of SUSY'01 for warmest hospitality and creating nice atmosphere during
the conference.
\vspace{-0.3cm}

\bibliographystyle{unsrt}

\begin{thebibliography}{99}


\bibitem{tevgr1}
\vspace{-0.2cm}
N. Arkani-Hamed, S. Dimopoulos, G. Dvali, Phys. Lett. B 429 (1998)
263; 
I. Antoniadis et al., Phys. Lett. B 436 (1998) 257.


\bibitem{bulknu1}
N. Arkani-Hamed et al., hep-ph/9811448.

\bibitem{neutsup}
E. Ma, M. Raidal, U. Sarkar,
hep-ph/0012101.

\bibitem{mypaper}
Z. Tavartkiladze, hep-ph/0105281.


\bibitem{osc}
%
R. Barbieri et al., hep-ph/9901228;
Q. Shafi, Z. Tavartkiladze, Phys. Lett. B 451 (1999) 129;
hep-ph/0101350; Phys. Lett. B 482 (2000) 145;
G. Altarelli, F. Feruglio, hep-ph/9905536;
see also references therein.

\bibitem{uni1}
K. Dines et al., hep-ph/9806292;
Z. Kakushadze, hep-th/9811193.



\end{thebibliography}

\end{document}